\documentclass[sigconf,screen]{acmart}
\AtBeginDocument{%
  \providecommand\BibTeX{{%
    \normalfont B\kern-0.5em{\scshape i\kern-0.25em b}\kern-0.8em\TeX}}}

\copyrightyear{2024} 
\acmYear{2024} 
\setcopyright{rightsretained} 
\acmConference[ICMI Companion '24]{INTERNATIONAL CONFERENCE ON
MULTIMODAL INTERACTION}{November 4--8, 2024}{San Jose, Costa Rica}
\acmBooktitle{INTERNATIONAL CONFERENCE ON MULTIMODAL INTERACTION
(ICMI Companion '24), November 4--8, 2024, San Jose, Costa
Rica}\acmDOI{10.1145/3686215.3690156}
\acmISBN{979-8-4007-0463-5/24/11}

\makeatletter
\gdef\@copyrightpermission{
  \begin{minipage}{0.3\columnwidth}
   \href{https://creativecommons.org/licenses/by-nd/4.0/}{\includegraphics[width=0.90\textwidth]{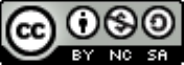}}
   \end{minipage}\hfill
   \begin{minipage}{0.7\columnwidth}
\href{https://creativecommons.org/licenses/by-nd/4.0/}{This work is licensed under a Creative Commons Attribution-NoDerivs International 4.0 License.}
  \end{minipage}
  \vspace{5pt}
}
\makeatother

\usepackage{xspace}
\usepackage{soul}
\usepackage{upgreek}
\usepackage{euscript}
\usepackage{booktabs}
\usepackage{siunitx}

\usepackage{adjustbox}
\usepackage{flushend}
\usepackage{balance}
\usepackage{subcaption} 

\usepackage{multirow}

\usepackage[normalem]{ulem}

\definecolor{forestgreen(web)}{rgb}{0.13, 0.55, 0.13}

\newcommand{\rND}{{\small \sf rND@4}\xspace}
\newcommand{\rKL}{{\small \sf rKL@4}\xspace}
\newcommand{\rRD}{{\small \sf rRD@4}\xspace}

\newcommand{\pro}{{\small \sf PRO}\xspace}
\newcommand{\con}{{\small \sf CON}\xspace}

\newcommand{\summarizer}{{\small \sf ChatGPT}\xspace}

\newcommand{\eg}{e.g.,\xspace}
\newcommand{\ie}{i.e.,\xspace}

\newcommand{\serp}{{\small \sf SERP}\xspace}
\newcommand{\seme}{{\small \sf SEME}\xspace}
\newcommand{\scs}{{\small \sf SCS}\xspace}

\newcommand{\ofair}{{$O_{\text{fair}}$}\xspace}
\newcommand{\efair}{{$E_{\text{fair}}$}\xspace}
\newcommand{\ounfair}{{$O_{\text{unfair}}$}\xspace}
\newcommand{\eunfair}{{$E_{\text{unfair}}$}\xspace}

\begin{document}

\title{Towards Investigating Biases in Spoken Conversational Search}


\author{Sachin Pathiyan Cherumanal}
\orcid{https://orcid.org/0000-0001-9982-3944} 
\affiliation{
\institution{RMIT University}
\city{Melbourne}
\country{Australia}
}
\email{s3874326@student.rmit.edu.au}

\author{Falk Scholer}
\orcid{https://orcid.org/0000-0001-9094-0810} 
\affiliation{
\institution{RMIT University}
\city{Melbourne}
\country{Australia}
}
\email{falk.scholer@rmit.edu.au}

\author{Johanne R.~Trippas}
\orcid{https://orcid.org/0000-0002-7801-0239} 
\affiliation{
\institution{RMIT University}
\city{Melbourne}
\country{Australia}
}
\email{j.trippas@rmit.edu.au}

\author{Damiano Spina}
\orcid{https://orcid.org/0000-0001-9913-433X}
\affiliation{
\institution{RMIT University}
\city{Melbourne}
\country{Australia}
}
\email{damiano.spina@rmit.edu.au}


\begin{abstract}

Voice-based systems like Amazon Alexa, Google Assistant, and Apple Siri, along with the growing popularity of OpenAI's ChatGPT and Microsoft's Copilot, serve diverse populations, including visually impaired and low-literacy communities. This reflects a shift in user expectations from traditional search to more interactive question-answering models. However, presenting information effectively in voice-only channels remains challenging due to their linear nature. This limitation can impact the presentation of complex queries involving controversial topics with multiple perspectives. Failing to present diverse viewpoints may perpetuate or introduce biases and affect user attitudes. Balancing information load and addressing biases is crucial in designing a fair and effective voice-based system. To address this, we 
\textit{(i)} review how biases and user attitude changes have been studied in screen-based web search, 
\textit{(ii)} address challenges in studying these changes in voice-based settings like SCS, 
\textit{(iii)} outline research questions, and 
\textit{(iv)} propose an experimental setup with variables, data, and instruments to explore biases in a voice-based setting like Spoken Conversational Search.
\end{abstract}



\maketitle

\section{Introduction}

The adoption of generative AI in search engines such as OpenAI's ChatGPT and Microsoft Copilot signifies a shift in traditional information retrieval. This transition moves away from the conventional ``ten blue links'' approach to a more conversational paradigm. While current generative AI technologies are mostly text-based, the increasing adoption of intelligent assistants like Google Assistant, Amazon Alexa, and Apple Siri indicates a move towards fully voice-based information interaction. Such Spoken Conversational Search (\scs) -- a type of voice-based information access system, can meet diverse information needs for many, including visually impaired, low-literacy, and sighted users in situations where reading is impractical, like driving, cooking, or exercising. 

Information access systems, like search engines, have traditionally been users' main source of information for decades. However, voice-based systems like \scs, and multimodal systems that integrate voice, face challenges due to the transient and linear nature of the voice channel, which can limit the amount of information that can be presented. Unlike screen-based web search that comes with the luxury of visually inspecting the ``ten-blue links'' anytime during the search session. For instance, in an \scs query about controversial topics like ``should zoos exist?'', the system should provide both sides (\ie~\pro and \con) to avoid biasing the user with one-sided perspectives. On a larger scale, this can adversely affect society by creating or reinforcing biases affecting users' attitudes towards certain topics. Such biased exposure of perspectives and its impact have been extensively studied in screen-based web search, but less in voice-based settings like \scs. 
In this work, we review attitude changes in screen-based web search (Section~\ref{sec:bias_IA}), address challenges in studying these changes in voice-based settings like \scs (Section~\ref{sec:bias_voice}), outline research questions (Section~\ref{sec:rqs}), and propose an experimental setup (Section~\ref{sec:exp_setup}) with variables, data, and instruments to explore biases in \scs.
\section{Background}

\subsection{Biases in Information Access}
\label{sec:bias_IA}
Search systems have become a primary source of information, but the resulting information overload~\cite{yunkaporta2023right} due to the vast amount of information available online, can cause stress, confusion, and reduce productivity~\cite{ji2024towards}. 
Effectively navigating information requires filtering, prioritizing, and organizing of the information. Due to limited cognitive capacity, users often rely on mental shortcuts, known as cognitive biases, which is a ``pattern of deviation'' from norm or rationality in thinking and reasoning, that influences the users' decision making process. For instance, during a search session, a user's attitude or decision can be heavily influenced by the order of the documents in the Search Engine Results Page (\serp) \ie \textit{order effect}, primarily due to position bias~\cite{pan2007google} -- a user's tendency to interact more with items at the top of a \serp.  A users decision making may also be influenced by the balanced or imbalanced exposure of perspectives \ie \textit{exposure effect}. For instance, \citet{gao2020toward} demonstrates a Google search query ``coffee health'' where the \serp fails to provide a balanced perspective on the benefits and the negative impacts of coffee on health. The two effects (\ie~ order and exposure) are two of many phenomenon that leads to change in user attitude is called Search Engine Manipulation Effect (\seme). Biases have also shown to be easily ``weaponized''~\cite{alaofi2022queries} to manipulate public opinion. For instance, ~\citet{epstein2015SEMEelections} shows that biased search rankings can shift the voting preferences of undecided voters by 20\% consequently impacting election outcomes.
Over the past years, \seme and biases in rankings have been extensively studied in the context of traditional screen-based web search \ie ``the ten blue links''. A recent crowdsourced study investigated factors affecting user attitude change while interacting with the entire \serp, including exposure and order effects, perceived diversity, pre-existing beliefs, and willingness to process counter-attitudinal information~\cite{draws21exploringbias}. \citet{bink2023snippetsuserattitudes} examined similar factors using \serp ``snippets''. However, investigating these factors in a voice-based setting like \scs remains an open challenge, as discussed in Section~\ref{sec:bias_voice}. To the best of our knowledge, the only biases studied in voice-setting are related to the recognition quality of phrases spoken by different groups~\cite{lima2019biasvoiceassistant}.

\subsection{Challenges in Spoken Conversational Search}
\label{sec:bias_voice}
Spoken Conversational Search (\scs) is an information access system where communication is entirely verbal~\cite{trippas2018scsperspectives}. The voice channel poses cognitive challenges, leading to different user behavior compared to a screen-based search processes. For instance, in traditional web search, users have the flexibility of reviewing multiple documents in the \serp with ten-blue links~\cite{zamani2023conversational, vtyurina2020mixed}. In contrast, in voice-only settings, the linear and transient nature of the channel may challenge users to keep up with information due to limited cognitive capacity. Another challenge is that users can easily review and refer back to their query with screens, which is inherently more difficult with voice queries. Moreover, previous work shows that users pay more attention to the first and last responses of a voice-based system~\cite{trippas2015resultspresentation} for faceted queries (e.g., broad intent with subtopics, like controversial queries). We are still unsure if the order of response influence user attitudes in a voice-based setting. Furthermore, when multiple perspectives need to be presented in an unbiased, it is important to understand if response order (\ie order effects) our source influences user attitude changes~\cite{trippas2024reevaluating}. One of the solutions to mitigate biases in screen-based web search has been to expose users to diverse perspectives. However, this becomes challenging in a voice-based setting due to user's limited cognitive capacity and the need to keep responses short to avoid additional cognitive load. Additionally, previous research indicates that user attitudes can change based on the proportion of perspectives consumed (\ie exposure effects, as noted by \citet{draws21exploringbias}). Following the same, we also explore how users' perceived diversity of voice responses and their open-mindedness impact their attitudes. With this knowledge, we define our research questions in Section \ref{sec:rqs}.
\section{Research Questions}
\label{sec:rqs}
The following research questions shall guide the experimental setup and protocol that will be used for our user study. 

\textbf{RQ1: \ }Does the order of passages (corresponding to different stances) affect user attitude in a voice based search system like \scs?

\textbf{RQ2: \ }Does the variation in exposure of passages (corresponding to different stances) affect user attitude in a voice based search system  like \scs?

\textbf{RQ3: \ }Do users actually perceive diversity in a voice-only search system? If so, is the ordering or the exposure of the passages perceived as diverse by the users of a voice based search system like \scs?

\textbf{RQ4: \ }Do factors like actively open-minded thinking (AOT) and perceived diversity help predict user attitude change in a voice-based search setting like \scs?
\section{Experimental Setup}
\label{sec:exp_setup}

In this section, we explore the potential experimental setup needed to address our research questions through a between-subjects 4x5 user study.\footnote{More details about the pre--registration are made available at~\url{https://osf.io/35ewj/}} 

\subsection{Data}
\label{subsec:data}

We use the annotated data released by \citet{draws21exploringbias}, which includes topics (\eg Is obesity a disease?), viewpoints (\ie~\pro or \con), and binary relevance (\ie relevant or not relevant). To control for unknown effects from non-relevant documents, we use only relevant ones. This dataset features five topics with neutral or mild attitudes: (i) \textit{Are social networking sites good for our society?}, (ii) \textit{Should zoos exist?}, (iii) \textit{Is cell phone radiation safe?}, (iv) \textit{Should bottled water be banned?}, (v) \textit{Is obesity a disease?}. While this data contains viewpoint labels in the range [$-3$,$3$], ($-3$ represents \textit{strongly opposing} and $+3$ represents \textit{strongly supporting}), for our study we pick arguments that have only the following labels $-2$, $-1$, $+1$ and $+2$ (due to lack of documents for the other viewpoints). For simplicity and to reduce the dimensions in our study, we group all the supporting passages and opposing passages to two different categories. The aforementioned documents are webpages, and not feasible in a voice-only setting due to audio limitations on search result presentation. Furthermore, previous work has shown that the length of voice responses can affect voice-interaction experiences~\cite{trippas2015length,trippas2015resultspresentation}. Section~\ref{subsubsec:document_passage_transform} details the steps we follow to control for the length of the passage.

\subsubsection{Passages}
\label{subsubsec:document_passage_transform}
We use \summarizer for aspect-based abstractive summarization~\cite{Lin_Ng_2019} of each document. Following \citet{Wu2022chainer}, we apply a zero-shot prompt chaining approach with the structure as follows. PROMPT 1:``\emph{From the text below, provide the justification, for [VIEWPOINT] the topic [TOPIC]. Justification: Text: [DOCUMENT CONTENT] }'',  where [VIEWPOINT] can be either ``attacking'' (\ie~\con) or ``supporting'' (\ie~\pro) and the [TOPIC] can be one of five discussed earlier (\eg \textit{Should zoos exist}). The output of PROMPT 1 is then chained as an input to PROMPT 2: ``\emph{Now summarize the below text in less than 30 words. Text: [PROMPT 1 RESPONSE]}''.

\subsubsection{Text-to-Speech}
\label{subsubsec:tts}
After generating the passages, we use Amazon Polly\footnote{\url{https://aws.amazon.com/polly/}} Text-To-Speech for synthesizing the voice responses, maintaining the same gender voice across all passages and avoiding audio manipulations (like pitch and pauses) to prevent impacting user information consumption \cite{chuklin2019using}. By using the same voice and tone features of the speech across all the passages, we ensure that there are no external factors (\eg persona, persuasiveness of the voice, gender) that impact the user \cite{persuasive2020Dubiel}.

\subsubsection{Voice-Only Search Results}
\label{subsubsec:search_results_conditions}
\begin{figure}[tp]
    \centering
    
    \begin{subfigure}[b]{1\linewidth}
        \centering
        \includegraphics[trim={0 9.2cm 0 5.5cm},clip,width=0.9\linewidth]{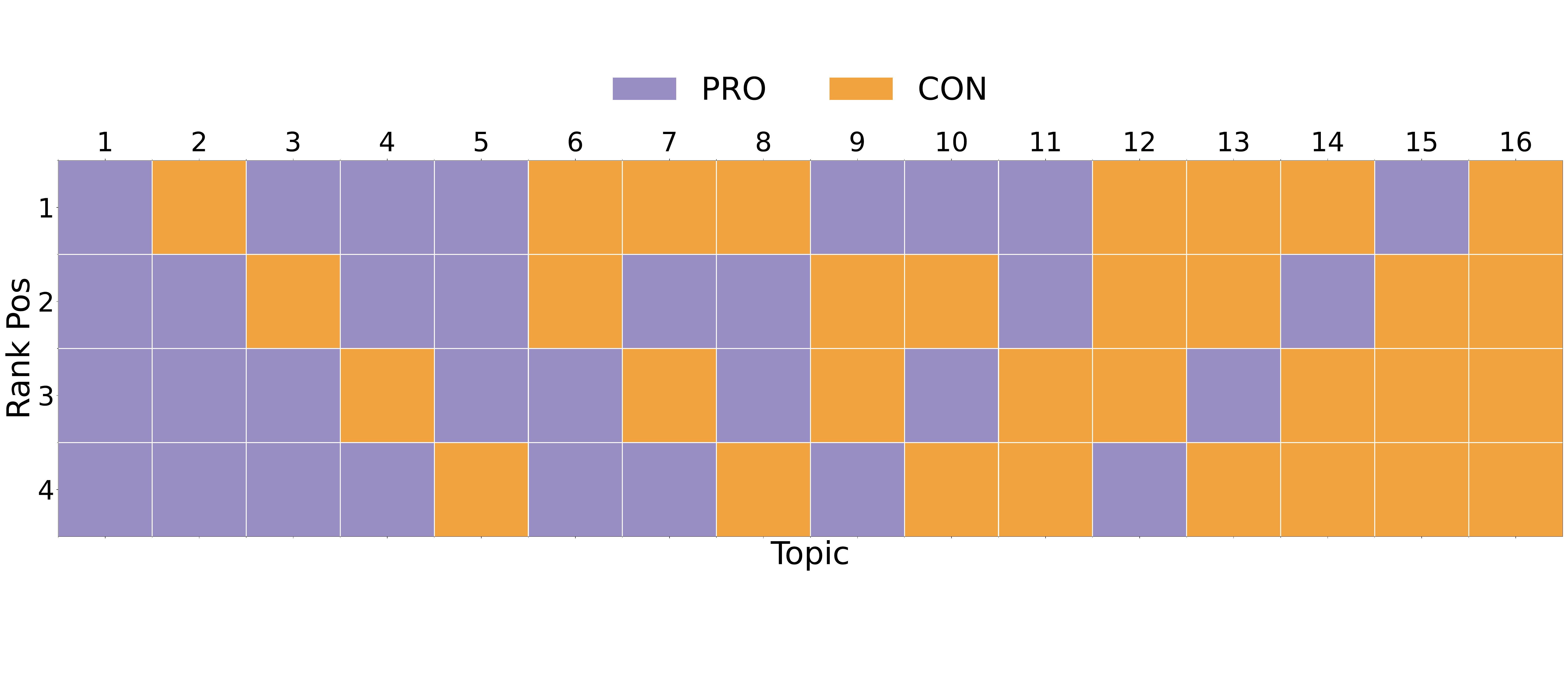}
        \Description[Illustration of permutations of stances across the top-4 positions using simulated scenarios.]{Illustration of Permutations of argument stances across the top-4 positions using simulated scenarios.}
        \caption{Permutations of argument stances across the top-4 positions using simulated scenarios.}
        \label{fig:synthetic_generation_matrix_top_4}
    \end{subfigure}
    
    \vspace{0.4cm} 

    \begin{subfigure}[b]{1\linewidth}
        \centering
        \includegraphics[width=0.9\linewidth]{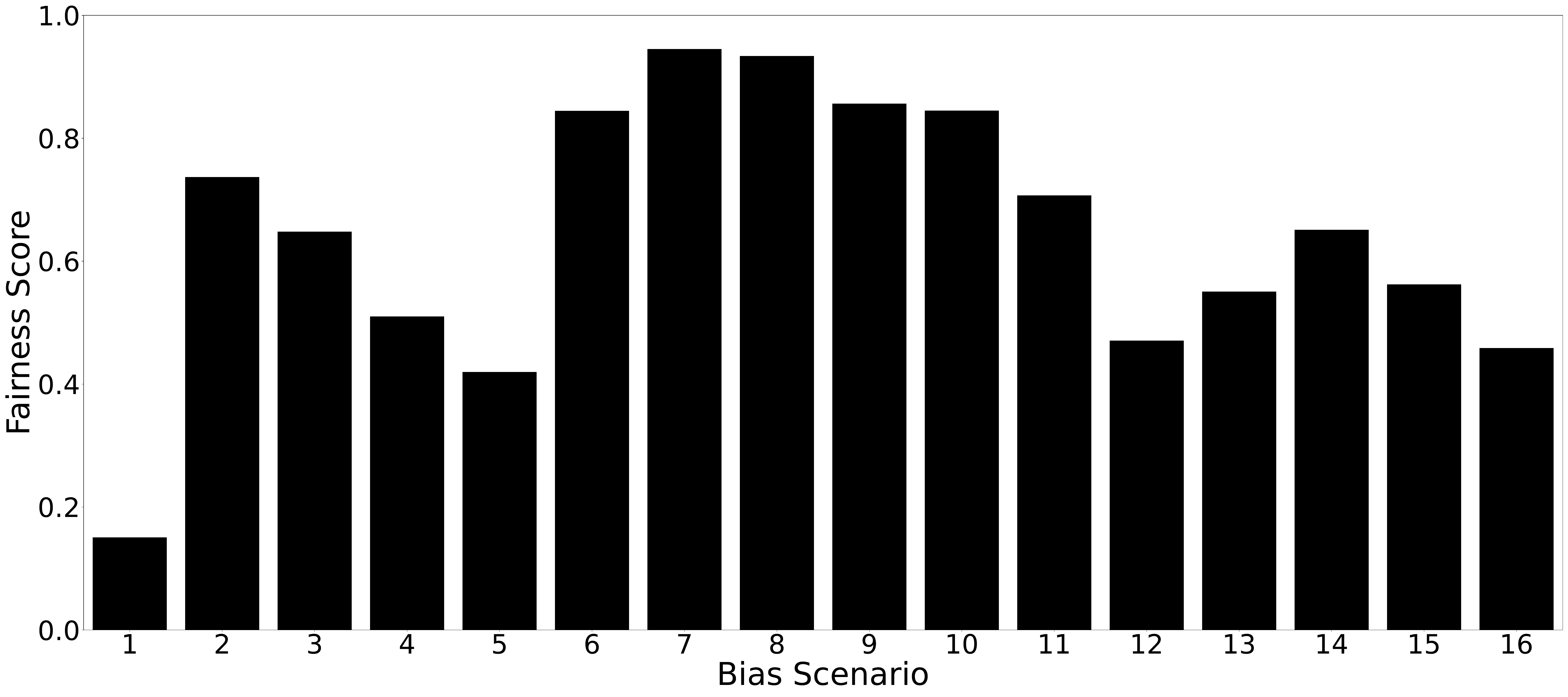}
        \Description[Illustration of Fairness score combined from three fairness metrics ~\cite{pathiyan2021evaluatingfairness}]{Illustration of Fairness score combined from three fairness metrics ~\cite{pathiyan2021evaluatingfairness}}
        \caption{Fairness score (higher means more fair) computed using three fairness metrics ~\cite{pathiyan2021evaluatingfairness}.}
        \label{fig:unfairness_combined_score}
    \end{subfigure}

    \caption{From \ref{fig:synthetic_generation_matrix_top_4} and \ref{fig:unfairness_combined_score}, we identify the $4$ topics/scenarios based on fairness score and narrow down out study to four scenarios (see Figure~\ref{fig:reduced_all_systems_order_exposure}).}
    \label{fig:combined_figure}
\end{figure}

We now use synthesized audio passages to create a ranked list containing top-$4$ results for each topic (as we focus on voice-only setting) in contrast to \citet{draws21exploringbias} where top-$10$ text-based results were used to simulate a web search experience. In Figure \ref{fig:synthetic_generation_matrix_top_4}, we show the $16$ possible permutations in which the two stances (\pro and \con) can be arranged. We also ensure that the synthetic scenario is proportion-agnostic, \ie all permutations have a $1/1$ ratio of \pro/\con arguments judged as relevant in the ground truth~\cite{pathiyan2021evaluatingfairness}. We then categorise the $16$ permutations into two (i) \textit{Exposure varying} ($E$), where the number of \pro and \con passages are not equal (ii) \textit{Exposure balanced} ($O$), where the number of \pro and \con passages in the top-$4$ position are balanced. Given the 16 ranked lists, we compute the fairness score using a combination of \rND, \rKL, \rRD adapted by ~\citet{pathiyan2021evaluatingfairness} specifically for conversational search on controversial topics (see Figure \ref{fig:unfairness_combined_score}). 

\begin{figure}[ht]
    \centering
    \includegraphics[trim={0 1.4cm 0 0.5cm},clip, width=0.45\linewidth]{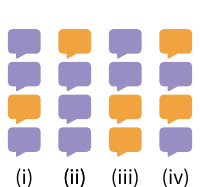}
    \Description[Illustration of the four conditions used in the study, (i) Exposure varying and Unfair (\eunfair) (ii) Exposure varying and fair (\efair) (iii) Exposure balanced and unfair (\ounfair) (iv) Exposure balanced and fair (\ofair). The four conditions represent top-$4$ positions in a ranking where each audio response has an associated stance/perspective \ie~ \pro (supporting) or \con (opposing).]{Illustration of the four conditions used in the study, (i) Exposure varying and Unfair (\eunfair) (ii) Exposure varying and fair (\efair) (iii) Exposure balanced and unfair (\ounfair) (iv) Exposure balanced and fair (\ofair). The four conditions represent top-$4$ positions in a ranking where each audio response has an associated stance/perspective \ie~ \pro (supporting) or \con (opposing).}
    \caption{The four conditions used in the study, (i) Exposure varying and Unfair (\eunfair) (ii) Exposure varying and fair (\efair) (iii) Exposure balanced and unfair (\ounfair) (iv) Exposure balanced and fair (\ofair). The four conditions represent top-$4$ positions in a ranking where each audio response has an associated stance/perspective \ie~ \pro (supporting) or \con (opposing).}
    \label{fig:reduced_all_systems_order_exposure}
    
\end{figure}

We pick one unfair and a fair system each of the two categories described above. In summary our study uses four systems (or conditions), (i) Exposure varying and Unfair (\eunfair) -- Permutation $4$ (ii) Exposure varying and fair (\efair) -- Permutation $2$ (iii) Exposure balanced and unfair (\ounfair) -- Permutation $11$ (iv) Exposure balanced and fair (\ofair) -- Permutation $7$ (see Figure \ref{fig:reduced_all_systems_order_exposure}). In  the case of \ofair and \ounfair the exposure is maintained in the top-4 positions. Whereas \efair 
 and \eunfair vary in terms of exposure. 

\begin{table*}[th]
\centering
\caption{Experiment variables considered in the study.}
\label{tab:variable-descriptions}
\adjustbox{max width=\textwidth}{
\begin{tabular}{p{1cm}p{4.1cm}p{11cm}}
\toprule
\multicolumn{1}{l}{\textbf{Variable   Type}} & \textbf{Variables}                  & \textbf{Description}                                                                                                                                                                                                                                                                                                                                                              \\
\midrule
\multirow{2}{*}{Independent}                 & Topic (x5)                               & Five topics the user will be   exposed to in the experiment.                                                                                                                                                                                                                                                                                                                      \\
                                             & Bias Scenario (x4)                       & One of the four search results   conditions discussed earlier in Section   \ref{subsubsec:search_results_conditions} and shown in Figure   \ref{fig:reduced_all_systems_order_exposure}.                                                                                                                                                                                          \\
                                             \midrule
Dependent                                    & User Attitude                       & Measured twice--before the experiment and after exposure to bias—- on a 7-point Likert Scale. The change will be calculated as in prior research in the web search context~\cite{draws21exploringbias}. \\
\midrule
\multirow{2}{*}{Covariates}                  & Actively Open-minded Thinking       & Measures the degree to which a person is willing to consider opposing perspectives-- on a 7-point Likert Scale~\cite{Haran_Ritov_Mellers_2013}.                                                                                                                                                                                                         \\
                                             & Perceived Diversity                 & Measures the degree of diversity in perspectives perceived by the user, in the voice-only search results-- on a 7-point Likert Scale~\cite{knijnenburg2012explaining}.                                                                                                                                                                                                                                \\
                                             \midrule
\multirow{5}{*}{Exploratory}                 & Gender and Age                             & \citet{martinez2023nobody}                                                                                                                                                                                                                                                                                                                                                        \\
                                             & Education Level                     & \citet{leory2019audioinformationsmartspeakers} found that education level impacts comprehension of information presented via audio channels.                                                                                                                                                                                                                                                      \\
                                             & Topic Familiarity                         & 3-point Likert Scale.                                                                                                                                                                                                                                                                                                                                                              \\
                                             & Declared Impairments                & Taking into account individual   variations for those with brain injuries or neurological conditions (e.g.,   neurodiverse population)~\cite{sitbon2023perspectives}.                                                                                                                                                                                                              \\
                                             & Language Proficiency                & \citet{muda2023people}     \\                                                                                                                                                                                                                                                                                                    \bottomrule  
\end{tabular}
}
\end{table*}

\subsection{Participants}
We propose to conduct a crowdsourced user study and identified the sample size required for a \textit{Between-Subjects Analysis of Covariance ANCOVA analysis} using the \textit{G* Power} software. Consequently, we propose to recruit $400$ participants from three English-speaking countries for our study (\ie USA, UK, and Australia). 

\subsection{Experiment Protocol}
\label{sec:protocol}

The recruited participants will be given backstories to simulate a real-life information need. The backstory will provide context for the voice-only search task to promote a natural search behaviour \cite{ghosh2019exploring}.
The participants will then be provided with a sample task to familiarise with the experiment setup. The topic used for this sample task shall be different from the five topics that will be used in this study. To ensure reproducibility of the web search context, we use the same backstory used by \citet{draws21exploringbias}. The steps involved in the task are mentioned below:

\paragraph{Step 1} Participants shall be given a brief overview of the task and asked to state their gender, age, familiarity with voice-only systems~\cite{kiesel2020expectations} and attitude towards one of the five debated topics on a 7-point Likert scale. The participant will be assigned the topic towards which they have a neutral attitude. For participants who does not have a neutral attitude, one of the 5 topics would be randomly assigned. Participants shall be assigned a topic toward which they have a mild pre-existing attitude (\ie~ responding with ``somewhat supporting'', ``neutral'', or ``somewhat opposing'').

\paragraph{Step 2} The participants shall then have the freedom to raise a spoken natural language query as this would help make the task more realistic. However, the query would not affect the results provided for the assigned topic.

\paragraph{Step 3} Participants will listen to four voice responses corresponding to the top four results from the ranked list for the given topic. However, participants shall be randomly assigned one of the bias scenarios shown in Figure \ref{fig:reduced_all_systems_order_exposure}.

\paragraph{Step 4} Participants would then state their (updated) attitude and interest concerning their assigned topic.

\paragraph{Step 5} The participants would be asked to fill a post-questionnaire that consisted of the AOT scale~\cite{Haran_Ritov_Mellers_2013} and the perceived diversity scale~\cite{knijnenburg2012explaining}.

\paragraph{Step 6} Participants will receive a written transcription of the voice responses they heard and will then be asked to re-order the passages according to the original sequence.
\section{Implications and Ethical Considerations}

Understanding how the order (\ie order effect) and proportion (\ie exposure effect) of perspectives in a voice-only setting influence user attitudes can inform the design of fairness-aware presentation strategies, an ongoing challenge~\cite{spina2024quantifying, spina2021futureconv}. Additionally, offline evaluation metrics for \scs and multimodal information access systems with voice interaction are still in their infancy, and often relies on web-search based user models (\eg logarithmic decay)~\cite{frummet2024report} that do not account for limitations in users' working memory and recall in voice channels. Our setup and recall tests may reveal patterns in user attention, provoking the community to explore \scs specific user models. These patterns could also inform and guide practitioners in designing techniques to alert users to potential biases, similar to the auditory warnings by ~\citet{pathiyan2024everything} and the ``nudges'' by ~\citet{Gohsen2023Nudge}. A limitation of our study is scalability due to the controlled knowledge base. However, advances in Large Language Models (LLMs), like Retrieval-Augmented Generation (RAG) or in general, Generative Information Access~\cite{trippas2024adapting}, could enable future experiments to use a broader knowledge base, as shown by~\citet{pathiyan2024walert}.

\paragraph{Ethical Considerations}
It may be worth keeping in mind that individual differences may exist between users, such as neurological impairments affecting comprehension. This must be considered~\cite{sitbon2023perspectives} in experiments that involves voice-based systems like \scs. A key challenge in systems like \scs is explainability~\cite{lajewska2024explainability} and the lack of explanations from \scs in such studies may impact users' trust and perceived credibility.
Furthermore, in voice-based settings exploring biases, participants are exposed to biased information and researchers may have to provide a textual summary afterward to prevent retention of biased content.
\section{Conclusion}
Given the impact of biases in traditional search and users' growing preference for conversational search experience powered by Large Language Models (LLMs), such as Microsoft Copilot and OpenAI’s {GPT--4o}\footnote{\url{https://openai.com/index/hello-gpt-4o/}} which prioritize natural language and voice, there is a critical need for research on biases in these settings due to a lack of exploration concerning biases in voice-based systems as discussed by \citet{ji2024towards,ji2024characterizingISprocesses}. This study focuses specifically on voice responses, aiming to investigate biases within this modality.


\begin{acks}
This research has been carried out in the unceded lands of the Woi Wurrung and Boon Wurrung
peoples of the eastern Kulin Nation. We pay our respects to their Ancestors and Elders, past,
present, and emerging. This research is partially supported by the Australian Research Council
(ARC, project nr. DE200100064 and CE200100005).
\end{acks}

\bibliographystyle{ACM-Reference-Format}
\bibliography{99-references}


\begin{thebibliography}{37}


\ifx \showCODEN    \undefined \def \showCODEN     #1{\unskip}     \fi
\ifx \showDOI      \undefined \def \showDOI       #1{#1}\fi
\ifx \showISBNx    \undefined \def \showISBNx     #1{\unskip}     \fi
\ifx \showISBNxiii \undefined \def \showISBNxiii  #1{\unskip}     \fi
\ifx \showISSN     \undefined \def \showISSN      #1{\unskip}     \fi
\ifx \showLCCN     \undefined \def \showLCCN      #1{\unskip}     \fi
\ifx \shownote     \undefined \def \shownote      #1{#1}          \fi
\ifx \showarticletitle \undefined \def \showarticletitle #1{#1}   \fi
\ifx \showURL      \undefined \def \showURL       {\relax}        \fi
\providecommand\bibfield[2]{#2}
\providecommand\bibinfo[2]{#2}
\providecommand\natexlab[1]{#1}
\providecommand\showeprint[2][]{arXiv:#2}

\bibitem[Alaofi et~al\mbox{.}(2022)]%
        {alaofi2022queries}
\bibfield{author}{\bibinfo{person}{Marwah Alaofi}, \bibinfo{person}{Luke Gallagher}, \bibinfo{person}{Dana Mckay}, \bibinfo{person}{Lauren~L. Saling}, \bibinfo{person}{Mark Sanderson}, \bibinfo{person}{Falk Scholer}, \bibinfo{person}{Damiano Spina}, {and} \bibinfo{person}{Ryen~W. White}.} \bibinfo{year}{2022}\natexlab{}.
\newblock \showarticletitle{Where Do Queries Come From?}. In \bibinfo{booktitle}{\emph{Proceedings of the 45th International ACM SIGIR Conference on Research and Development in Information Retrieval}} (Madrid, Spain) \emph{(\bibinfo{series}{SIGIR '22})}. \bibinfo{publisher}{Association for Computing Machinery}, \bibinfo{address}{New York, NY, USA}, \bibinfo{pages}{2850–2862}.
\newblock
\showISBNx{9781450387323}
\urldef\tempurl%
\url{https://doi.org/10.1145/3477495.3531711}
\showDOI{\tempurl}


\bibitem[Bink et~al\mbox{.}(2023)]%
        {bink2023snippetsuserattitudes}
\bibfield{author}{\bibinfo{person}{Markus Bink}, \bibinfo{person}{Sebastian Schwarz}, \bibinfo{person}{Tim Draws}, {and} \bibinfo{person}{David Elsweiler}.} \bibinfo{year}{2023}\natexlab{}.
\newblock \showarticletitle{Investigating the Influence of Featured Snippets on User Attitudes}. In \bibinfo{booktitle}{\emph{Proceedings of the 2023 Conference on Human Information Interaction and Retrieval}} (Austin, TX, USA) \emph{(\bibinfo{series}{CHIIR '23})}. \bibinfo{publisher}{Association for Computing Machinery}, \bibinfo{address}{New York, NY, USA}, \bibinfo{pages}{211–220}.
\newblock
\showISBNx{9798400700354}
\urldef\tempurl%
\url{https://doi.org/10.1145/3576840.3578323}
\showDOI{\tempurl}


\bibitem[Chuklin et~al\mbox{.}(2019)]%
        {chuklin2019using}
\bibfield{author}{\bibinfo{person}{Aleksandr Chuklin}, \bibinfo{person}{Aliaksei Severyn}, \bibinfo{person}{Johanne~R. Trippas}, \bibinfo{person}{Enrique Alfonseca}, \bibinfo{person}{Hanna Silen}, {and} \bibinfo{person}{Damiano Spina}.} \bibinfo{year}{2019}\natexlab{}.
\newblock \showarticletitle{Using Audio Transformations to Improve Comprehension in Voice Question Answering}. In \bibinfo{booktitle}{\emph{Experimental IR Meets Multilinguality, Multimodality, and Interaction: 10th International Conference of the CLEF Association, CLEF 2019, Lugano, Switzerland, September 9–12, 2019, Proceedings}} (Lugano, Switzerland). \bibinfo{publisher}{Springer-Verlag}, \bibinfo{address}{Berlin, Heidelberg}, \bibinfo{pages}{164–170}.
\newblock
\showISBNx{978-3-030-28576-0}
\urldef\tempurl%
\url{https://doi.org/10.1007/978-3-030-28577-7_12}
\showDOI{\tempurl}


\bibitem[Draws et~al\mbox{.}(2021)]%
        {draws21exploringbias}
\bibfield{author}{\bibinfo{person}{Tim Draws}, \bibinfo{person}{Nava Tintarev}, \bibinfo{person}{Ujwal Gadiraju}, \bibinfo{person}{Alessandro Bozzon}, {and} \bibinfo{person}{Benjamin Timmermans}.} \bibinfo{year}{2021}\natexlab{}.
\newblock \showarticletitle{This Is Not What We Ordered: Exploring Why Biased Search Result Rankings Affect User Attitudes on Debated Topics}. In \bibinfo{booktitle}{\emph{Proceedings of the 44th International ACM SIGIR Conference on Research and Development in Information Retrieval}} (Virtual Event, Canada) \emph{(\bibinfo{series}{SIGIR '21})}. \bibinfo{publisher}{Association for Computing Machinery}, \bibinfo{address}{New York, NY, USA}, \bibinfo{pages}{295–305}.
\newblock
\showISBNx{9781450380379}
\urldef\tempurl%
\url{https://doi.org/10.1145/3404835.3462851}
\showDOI{\tempurl}


\bibitem[Dubiel et~al\mbox{.}(2020)]%
        {persuasive2020Dubiel}
\bibfield{author}{\bibinfo{person}{Mateusz Dubiel}, \bibinfo{person}{Martin Halvey}, \bibinfo{person}{Pilar~Oplustil Gallegos}, {and} \bibinfo{person}{Simon King}.} \bibinfo{year}{2020}\natexlab{}.
\newblock \showarticletitle{Persuasive Synthetic Speech: Voice Perception and User Behaviour}. In \bibinfo{booktitle}{\emph{Proceedings of the 2nd Conference on Conversational User Interfaces}} (Bilbao, Spain) \emph{(\bibinfo{series}{CUI '20})}. \bibinfo{publisher}{Association for Computing Machinery}, \bibinfo{address}{New York, NY, USA}, Article \bibinfo{articleno}{6}, \bibinfo{numpages}{9}~pages.
\newblock
\showISBNx{9781450375443}
\urldef\tempurl%
\url{https://doi.org/10.1145/3405755.3406120}
\showDOI{\tempurl}


\bibitem[Epstein and Robertson(2015)]%
        {epstein2015SEMEelections}
\bibfield{author}{\bibinfo{person}{Robert Epstein} {and} \bibinfo{person}{Ronald~E. Robertson}.} \bibinfo{year}{2015}\natexlab{}.
\newblock \showarticletitle{The search engine manipulation effect (SEME) and its possible impact on the outcomes of elections}.
\newblock \bibinfo{journal}{\emph{Proceedings of the National Academy of Sciences}} \bibinfo{volume}{112}, \bibinfo{number}{33} (\bibinfo{year}{2015}), \bibinfo{pages}{E4512--E4521}.
\newblock
\urldef\tempurl%
\url{https://doi.org/10.1073/pnas.1419828112}
\showDOI{\tempurl}


\bibitem[Frummet et~al\mbox{.}(2024)]%
        {frummet2024report}
\bibfield{author}{\bibinfo{person}{Alexander Frummet}, \bibinfo{person}{Andrea Papenmeier}, \bibinfo{person}{Maik Fr\"{o}be}, \bibinfo{person}{Johannes Kiesel}, \bibinfo{person}{Vaibhav Adlakha}, \bibinfo{person}{Norbert Braunschweiler}, \bibinfo{person}{Mateusz Dubiel}, \bibinfo{person}{Satanu Ghosh}, \bibinfo{person}{Marcel Gohsen}, \bibinfo{person}{Christin Kreutz}, \bibinfo{person}{Milad Momeni}, \bibinfo{person}{Markus Nilles}, \bibinfo{person}{Sachin~Pathiyan Cherumanal}, \bibinfo{person}{Abbas Pirmoradi}, \bibinfo{person}{Paul Thomas}, \bibinfo{person}{Johanne~R. Trippas}, \bibinfo{person}{Ines Zelch}, {and} \bibinfo{person}{Oleg Zendel}.} \bibinfo{year}{2024}\natexlab{}.
\newblock \showarticletitle{Report on the 8th Workshop on Search-Oriented Conversational Artificial Intelligence (SCAI 2024) at CHIIR 2024}.
\newblock \bibinfo{journal}{\emph{SIGIR Forum}} \bibinfo{volume}{58}, \bibinfo{number}{1} (\bibinfo{date}{aug} \bibinfo{year}{2024}), \bibinfo{pages}{1–12}.
\newblock
\showISSN{0163-5840}
\urldef\tempurl%
\url{https://doi.org/10.1145/3687273.3687282}
\showDOI{\tempurl}


\bibitem[Gao and Shah(2020)]%
        {gao2020toward}
\bibfield{author}{\bibinfo{person}{Ruoyuan Gao} {and} \bibinfo{person}{Chirag Shah}.} \bibinfo{year}{2020}\natexlab{}.
\newblock \showarticletitle{Toward Creating a Fairer Ranking in Search Engine Results}.
\newblock \bibinfo{journal}{\emph{Information Processing \& Management}} \bibinfo{volume}{57}, \bibinfo{number}{1} (\bibinfo{year}{2020}), \bibinfo{pages}{102138}.
\newblock
\urldef\tempurl%
\url{https://doi.org/10.1016/j.ipm.2019.102138}
\showURL{%
\tempurl}


\bibitem[Ghosh(2019)]%
        {ghosh2019exploring}
\bibfield{author}{\bibinfo{person}{Souvick Ghosh}.} \bibinfo{year}{2019}\natexlab{}.
\newblock \showarticletitle{Exploring Result Presentation in Conversational IR Using a Wizard-of-Oz Study}. In \bibinfo{booktitle}{\emph{Advances in Information Retrieval: 41st European Conference on IR Research, ECIR 2019, Cologne, Germany, April 14–18, 2019, Proceedings, Part II}} (Cologne, Germany). \bibinfo{publisher}{Springer-Verlag}, \bibinfo{address}{Berlin, Heidelberg}, \bibinfo{pages}{327–331}.
\newblock
\showISBNx{978-3-030-15718-0}
\urldef\tempurl%
\url{https://doi.org/10.1007/978-3-030-15719-7_43}
\showDOI{\tempurl}


\bibitem[Gohsen et~al\mbox{.}(2023)]%
        {Gohsen2023Nudge}
\bibfield{author}{\bibinfo{person}{Marcel Gohsen}, \bibinfo{person}{Johannes Kiesel}, \bibinfo{person}{Mariam Korashi}, \bibinfo{person}{Jan Ehlers}, {and} \bibinfo{person}{Benno Stein}.} \bibinfo{year}{2023}\natexlab{}.
\newblock \showarticletitle{Guiding Oral Conversations: How to Nudge Users Towards Asking Questions?}. In \bibinfo{booktitle}{\emph{Proceedings of the 2023 Conference on Human Information Interaction and Retrieval}} (Austin, TX, USA) \emph{(\bibinfo{series}{CHIIR '23})}. \bibinfo{publisher}{Association for Computing Machinery}, \bibinfo{address}{New York, NY, USA}, \bibinfo{pages}{34–42}.
\newblock
\showISBNx{9798400700354}
\urldef\tempurl%
\url{https://doi.org/10.1145/3576840.3578291}
\showDOI{\tempurl}


\bibitem[Haran et~al\mbox{.}(2013)]%
        {Haran_Ritov_Mellers_2013}
\bibfield{author}{\bibinfo{person}{Uriel Haran}, \bibinfo{person}{Ilana Ritov}, {and} \bibinfo{person}{Barbara~A. Mellers}.} \bibinfo{year}{2013}\natexlab{}.
\newblock \showarticletitle{The role of actively open-minded thinking in information acquisition, accuracy, and calibration}.
\newblock \bibinfo{journal}{\emph{Judgment and Decision Making}} \bibinfo{volume}{8}, \bibinfo{number}{3} (\bibinfo{year}{2013}), \bibinfo{pages}{188–201}.
\newblock
\urldef\tempurl%
\url{https://doi.org/10.1017/S1930297500005921}
\showDOI{\tempurl}


\bibitem[Ji et~al\mbox{.}(2024a)]%
        {ji2024characterizingISprocesses}
\bibfield{author}{\bibinfo{person}{Kaixin Ji}, \bibinfo{person}{Danula Hettiachchi}, \bibinfo{person}{Flora~D. Salim}, \bibinfo{person}{Falk Scholer}, {and} \bibinfo{person}{Damiano Spina}.} \bibinfo{year}{2024}\natexlab{a}.
\newblock \showarticletitle{Characterizing Information Seeking Processes with Multiple Physiological Signals}. In \bibinfo{booktitle}{\emph{Proceedings of the 47th International ACM SIGIR Conference on Research and Development in Information Retrieval}} (Washington DC, USA) \emph{(\bibinfo{series}{SIGIR '24})}. \bibinfo{publisher}{Association for Computing Machinery}, \bibinfo{address}{New York, NY, USA}, \bibinfo{pages}{1006–1017}.
\newblock
\showISBNx{9798400704314}
\urldef\tempurl%
\url{https://doi.org/10.1145/3626772.3657793}
\showDOI{\tempurl}


\bibitem[Ji et~al\mbox{.}(2024b)]%
        {ji2024towards}
\bibfield{author}{\bibinfo{person}{Kaixin Ji}, \bibinfo{person}{Sachin Pathiyan~Cherumanal}, \bibinfo{person}{Johanne~R. Trippas}, \bibinfo{person}{Danula Hettiachchi}, \bibinfo{person}{Flora~D. Salim}, \bibinfo{person}{Falk Scholer}, {and} \bibinfo{person}{Damiano Spina}.} \bibinfo{year}{2024}\natexlab{b}.
\newblock \showarticletitle{Towards Detecting and Mitigating Cognitive Bias in Spoken Conversational Search}. In \bibinfo{booktitle}{\emph{Adjunct Publication of the 26th International Conference on Human-Computer Interaction with Mobile Devices and Services}} \emph{(\bibinfo{series}{MobileHCI '24})}. \bibinfo{publisher}{Association for Computing Machinery}, \bibinfo{address}{New York, NY, USA}.
\newblock
\urldef\tempurl%
\url{https://doi.org/10.1145/3640471.3680245}
\showDOI{\tempurl}


\bibitem[Kiesel et~al\mbox{.}(2020)]%
        {kiesel2020expectations}
\bibfield{author}{\bibinfo{person}{Johannes Kiesel}, \bibinfo{person}{Kevin Lang}, \bibinfo{person}{Henning Wachsmuth}, \bibinfo{person}{Eva Hornecker}, {and} \bibinfo{person}{Benno Stein}.} \bibinfo{year}{2020}\natexlab{}.
\newblock \showarticletitle{Investigating Expectations for Voice-Based and Conversational Argument Search on the Web}. In \bibinfo{booktitle}{\emph{Proceedings of the 2020 Conference on Human Information Interaction and Retrieval}} (Vancouver BC, Canada) \emph{(\bibinfo{series}{CHIIR '20})}. \bibinfo{publisher}{Association for Computing Machinery}, \bibinfo{address}{New York, NY, USA}, \bibinfo{pages}{53–62}.
\newblock
\showISBNx{9781450368926}
\urldef\tempurl%
\url{https://doi.org/10.1145/3343413.3377978}
\showDOI{\tempurl}


\bibitem[Knijnenburg et~al\mbox{.}(2012)]%
        {knijnenburg2012explaining}
\bibfield{author}{\bibinfo{person}{Bart~P Knijnenburg}, \bibinfo{person}{Martijn~C Willemsen}, \bibinfo{person}{Zeno Gantner}, \bibinfo{person}{Hakan Soncu}, {and} \bibinfo{person}{Chris Newell}.} \bibinfo{year}{2012}\natexlab{}.
\newblock \showarticletitle{Explaining the user experience of recommender systems}.
\newblock \bibinfo{journal}{\emph{User modeling and user-adapted interaction}}  \bibinfo{volume}{22} (\bibinfo{year}{2012}), \bibinfo{pages}{441--504}.
\newblock
\urldef\tempurl%
\url{https://doi.org/10.1007/s11257-011-9118-4}
\showDOI{\tempurl}


\bibitem[\L{}ajewska et~al\mbox{.}(2024)]%
        {lajewska2024explainability}
\bibfield{author}{\bibinfo{person}{Weronika \L{}ajewska}, \bibinfo{person}{Damiano Spina}, \bibinfo{person}{Johanne Trippas}, {and} \bibinfo{person}{Krisztian Balog}.} \bibinfo{year}{2024}\natexlab{}.
\newblock \showarticletitle{Explainability for Transparent Conversational Information-Seeking}. In \bibinfo{booktitle}{\emph{Proceedings of the 47th International ACM SIGIR Conference on Research and Development in Information Retrieval}} (Washington DC, USA) \emph{(\bibinfo{series}{SIGIR '24})}. \bibinfo{publisher}{Association for Computing Machinery}, \bibinfo{address}{New York, NY, USA}, \bibinfo{pages}{1040–1050}.
\newblock
\showISBNx{9798400704314}
\urldef\tempurl%
\url{https://doi.org/10.1145/3626772.3657768}
\showDOI{\tempurl}


\bibitem[Leroy and Kauchak(2019)]%
        {leory2019audioinformationsmartspeakers}
\bibfield{author}{\bibinfo{person}{Gondy Leroy} {and} \bibinfo{person}{David Kauchak}.} \bibinfo{year}{2019}\natexlab{}.
\newblock \showarticletitle{{A comparison of text versus audio for information comprehension with future uses for smart speakers}}.
\newblock \bibinfo{journal}{\emph{JAMIA Open}} \bibinfo{volume}{2}, \bibinfo{number}{2} (\bibinfo{date}{05} \bibinfo{year}{2019}), \bibinfo{pages}{254--260}.
\newblock
\showISSN{2574-2531}
\urldef\tempurl%
\url{https://doi.org/10.1093/jamiaopen/ooz011}
\showDOI{\tempurl}


\bibitem[Lima et~al\mbox{.}(2019)]%
        {lima2019biasvoiceassistant}
\bibfield{author}{\bibinfo{person}{Lanna Lima}, \bibinfo{person}{Vasco Furtado}, \bibinfo{person}{Elizabeth Furtado}, {and} \bibinfo{person}{Virgilio Almeida}.} \bibinfo{year}{2019}\natexlab{}.
\newblock \showarticletitle{Empirical Analysis of Bias in Voice-based Personal Assistants}. In \bibinfo{booktitle}{\emph{Companion Proceedings of The 2019 World Wide Web Conference}} (San Francisco, USA) \emph{(\bibinfo{series}{WWW '19})}. \bibinfo{publisher}{Association for Computing Machinery}, \bibinfo{address}{New York, NY, USA}, \bibinfo{pages}{533–538}.
\newblock
\showISBNx{9781450366755}
\urldef\tempurl%
\url{https://doi.org/10.1145/3308560.3317597}
\showDOI{\tempurl}


\bibitem[Lin and Ng(2019)]%
        {Lin_Ng_2019}
\bibfield{author}{\bibinfo{person}{Hui Lin} {and} \bibinfo{person}{Vincent Ng}.} \bibinfo{year}{2019}\natexlab{}.
\newblock \showarticletitle{Abstractive Summarization: A Survey of the State of the Art}.
\newblock \bibinfo{journal}{\emph{Proceedings of the AAAI Conference on Artificial Intelligence}} \bibinfo{volume}{33}, \bibinfo{number}{01} (\bibinfo{date}{Jul.} \bibinfo{year}{2019}), \bibinfo{pages}{9815--9822}.
\newblock
\urldef\tempurl%
\url{https://doi.org/10.1609/aaai.v33i01.33019815}
\showDOI{\tempurl}


\bibitem[Mart{\'\i}nez-Costa et~al\mbox{.}(2023)]%
        {martinez2023nobody}
\bibfield{author}{\bibinfo{person}{Mar{\'\i}a-Pilar Mart{\'\i}nez-Costa}, \bibinfo{person}{Fernando L{\'o}pez-Pan}, \bibinfo{person}{Nataly Busl{\'o}n}, {and} \bibinfo{person}{Ram{\'o}n Salaverr{\'\i}a}.} \bibinfo{year}{2023}\natexlab{}.
\newblock \showarticletitle{Nobody-fools-me perception: Influence of Age and Education on Overconfidence About Spotting Disinformation}.
\newblock \bibinfo{journal}{\emph{Journalism Practice}} \bibinfo{volume}{17}, \bibinfo{number}{10} (\bibinfo{year}{2023}), \bibinfo{pages}{2084--2102}.
\newblock
\urldef\tempurl%
\url{https://doi.org/10.1080/17512786.2022.2135128}
\showDOI{\tempurl}


\bibitem[Muda et~al\mbox{.}(2023)]%
        {muda2023people}
\bibfield{author}{\bibinfo{person}{Rafa{\l} Muda}, \bibinfo{person}{Gordon Pennycook}, \bibinfo{person}{Damian Hamerski}, {and} \bibinfo{person}{Micha{\l} Bia{\l}ek}.} \bibinfo{year}{2023}\natexlab{}.
\newblock \showarticletitle{People are worse at detecting fake news in their foreign language.}
\newblock \bibinfo{journal}{\emph{Journal of Experimental Psychology: Applied}} (\bibinfo{year}{2023}).
\newblock
\urldef\tempurl%
\url{https://doi.org/10.1037/xap0000475}
\showDOI{\tempurl}


\bibitem[Pan et~al\mbox{.}(2007)]%
        {pan2007google}
\bibfield{author}{\bibinfo{person}{Bing Pan}, \bibinfo{person}{Helene Hembrooke}, \bibinfo{person}{Thorsten Joachims}, \bibinfo{person}{Lori Lorigo}, \bibinfo{person}{Geri Gay}, {and} \bibinfo{person}{Laura Granka}.} \bibinfo{year}{2007}\natexlab{}.
\newblock \showarticletitle{{In Google We Trust: Users’ Decisions on Rank, Position, and Relevance}}.
\newblock \bibinfo{journal}{\emph{Journal of Computer-Mediated Communication}} \bibinfo{volume}{12}, \bibinfo{number}{3} (\bibinfo{date}{04} \bibinfo{year}{2007}), \bibinfo{pages}{801--823}.
\newblock
\showISSN{1083-6101}
\urldef\tempurl%
\url{https://doi.org/10.1111/j.1083-6101.2007.00351.x}
\showDOI{\tempurl}


\bibitem[Pathiyan~Cherumanal et~al\mbox{.}(2024a)]%
        {pathiyan2024everything}
\bibfield{author}{\bibinfo{person}{Sachin Pathiyan~Cherumanal}, \bibinfo{person}{Ujwal Gadiraju}, {and} \bibinfo{person}{Damiano Spina}.} \bibinfo{year}{2024}\natexlab{a}.
\newblock \showarticletitle{Everything We Hear: Towards Tackling Misinformation in Podcasts}. In \bibinfo{booktitle}{\emph{Proceedings of the 25th International Conference on Multimodal Interaction}} \emph{(\bibinfo{series}{ICMI '24})}. \bibinfo{publisher}{Association for Computing Machinery}, \bibinfo{address}{New York, NY, USA}.
\newblock
\urldef\tempurl%
\url{https://doi.org/10.1145/3678957.3678959}
\showDOI{\tempurl}


\bibitem[Pathiyan~Cherumanal et~al\mbox{.}(2021)]%
        {pathiyan2021evaluatingfairness}
\bibfield{author}{\bibinfo{person}{Sachin Pathiyan~Cherumanal}, \bibinfo{person}{Damiano Spina}, \bibinfo{person}{Falk Scholer}, {and} \bibinfo{person}{W.~Bruce Croft}.} \bibinfo{year}{2021}\natexlab{}.
\newblock \showarticletitle{Evaluating Fairness in Argument Retrieval}. In \bibinfo{booktitle}{\emph{Proceedings of the 30th ACM International Conference on Information \& Knowledge Management}} (Virtual Event, Queensland, Australia) \emph{(\bibinfo{series}{CIKM '21})}. \bibinfo{publisher}{Association for Computing Machinery}, \bibinfo{address}{New York, NY, USA}, \bibinfo{pages}{3363–3367}.
\newblock
\showISBNx{9781450384469}
\urldef\tempurl%
\url{https://doi.org/10.1145/3459637.3482099}
\showDOI{\tempurl}


\bibitem[Pathiyan~Cherumanal et~al\mbox{.}(2024b)]%
        {pathiyan2024walert}
\bibfield{author}{\bibinfo{person}{Sachin Pathiyan~Cherumanal}, \bibinfo{person}{Lin Tian}, \bibinfo{person}{Futoon~M. Abushaqra}, \bibinfo{person}{Angel~Felipe Magnoss\~{a}o~de Paula}, \bibinfo{person}{Kaixin Ji}, \bibinfo{person}{Halil Ali}, \bibinfo{person}{Danula Hettiachchi}, \bibinfo{person}{Johanne~R. Trippas}, \bibinfo{person}{Falk Scholer}, {and} \bibinfo{person}{Damiano Spina}.} \bibinfo{year}{2024}\natexlab{b}.
\newblock \showarticletitle{Walert: Putting Conversational Information Seeking Knowledge into Action by Building and Evaluating a Large Language Model-Powered Chatbot}. In \bibinfo{booktitle}{\emph{Proceedings of the 2024 Conference on Human Information Interaction and Retrieval}} (Sheffield, United Kingdom) \emph{(\bibinfo{series}{CHIIR '24})}. \bibinfo{publisher}{Association for Computing Machinery}, \bibinfo{address}{New York, NY, USA}, \bibinfo{pages}{401–405}.
\newblock
\showISBNx{9798400704345}
\urldef\tempurl%
\url{https://doi.org/10.1145/3627508.3638309}
\showDOI{\tempurl}


\bibitem[Sitbon et~al\mbox{.}(2023)]%
        {sitbon2023perspectives}
\bibfield{author}{\bibinfo{person}{Laurianne Sitbon}, \bibinfo{person}{Gerd Berget}, {and} \bibinfo{person}{Margot Brereton}.} \bibinfo{year}{2023}\natexlab{}.
\newblock \showarticletitle{Perspectives of Neurodiverse Participants in Interactive Information Retrieval}.
\newblock \bibinfo{journal}{\emph{Foundations and Trends{\textregistered} in Information Retrieval}} \bibinfo{volume}{17}, \bibinfo{number}{2} (\bibinfo{year}{2023}), \bibinfo{pages}{124--243}.
\newblock
\urldef\tempurl%
\url{https://doi.org/10.1561/1500000086}
\showDOI{\tempurl}


\bibitem[Spina et~al\mbox{.}(2024)]%
        {spina2024quantifying}
\bibfield{author}{\bibinfo{person}{Damiano Spina}, \bibinfo{person}{Danula Hettiachchi}, {and} \bibinfo{person}{Anthony McCosker}.} \bibinfo{year}{2024}\natexlab{}.
\newblock \bibinfo{booktitle}{\emph{Quantifying and Measuring Bias and Engagement in Automated Decision-Making}}.
\newblock \bibinfo{type}{{T}echnical {R}eport}. \bibinfo{institution}{ARC Centre of Excellence for Automated Decision-Making and Society, RMIT University}, \bibinfo{address}{Melbourne, Australia}.
\newblock
\urldef\tempurl%
\url{https://doi.org/10.60836/k8xh-en92}
\showDOI{\tempurl}


\bibitem[Spina et~al\mbox{.}(2021)]%
        {spina2021futureconv}
\bibfield{author}{\bibinfo{person}{Damiano Spina}, \bibinfo{person}{Johanne~R. Trippas}, \bibinfo{person}{Paul Thomas}, \bibinfo{person}{Hideo Joho}, \bibinfo{person}{Katriina Bystr\"{o}m}, \bibinfo{person}{Leigh Clark}, \bibinfo{person}{Nick Craswell}, \bibinfo{person}{Mary Czerwinski}, \bibinfo{person}{David Elsweiler}, \bibinfo{person}{Alexander Frummet}, \bibinfo{person}{Souvick Ghosh}, \bibinfo{person}{Johannes Kiesel}, \bibinfo{person}{Irene Lopatovska}, \bibinfo{person}{Daniel McDuff}, \bibinfo{person}{Selina Meyer}, \bibinfo{person}{Ahmed Mourad}, \bibinfo{person}{Paul Owoicho}, \bibinfo{person}{Sachin~Pathiyan Cherumanal}, \bibinfo{person}{Daniel Russell}, {and} \bibinfo{person}{Laurianne Sitbon}.} \bibinfo{year}{2021}\natexlab{}.
\newblock \showarticletitle{Report on the future conversations workshop at CHIIR 2021}.
\newblock \bibinfo{journal}{\emph{SIGIR Forum}} \bibinfo{volume}{55}, \bibinfo{number}{1}, Article \bibinfo{articleno}{6} (\bibinfo{date}{jul} \bibinfo{year}{2021}), \bibinfo{numpages}{22}~pages.
\newblock
\showISSN{0163-5840}
\urldef\tempurl%
\url{https://doi.org/10.1145/3476415.3476421}
\showDOI{\tempurl}


\bibitem[Trippas et~al\mbox{.}(2024a)]%
        {trippas2024reevaluating}
\bibfield{author}{\bibinfo{person}{Johanne~R Trippas}, \bibinfo{person}{Luke Gallagher}, {and} \bibinfo{person}{Joel Mackenzie}.} \bibinfo{year}{2024}\natexlab{a}.
\newblock \showarticletitle{Re-evaluating the Command-and-Control Paradigm in Conversational Search Interactions}. In \bibinfo{booktitle}{\emph{Proceedings of the 33rd ACM International Conference on Information and Knowledge Management}} (Boise, ID, USA) \emph{(\bibinfo{series}{CIKM '24})}. \bibinfo{publisher}{ACM}, \bibinfo{address}{New York, NY, USA}.
\newblock
\urldef\tempurl%
\url{https://doi.org/doi.org/10.1145/3627673.3679588}
\showDOI{\tempurl}


\bibitem[Trippas et~al\mbox{.}(2018)]%
        {trippas2018scsperspectives}
\bibfield{author}{\bibinfo{person}{Johanne~R. Trippas}, \bibinfo{person}{Damiano Spina}, \bibinfo{person}{Lawrence Cavedon}, \bibinfo{person}{Hideo Joho}, {and} \bibinfo{person}{Mark Sanderson}.} \bibinfo{year}{2018}\natexlab{}.
\newblock \showarticletitle{Informing the Design of Spoken Conversational Search: Perspective Paper}. In \bibinfo{booktitle}{\emph{Proceedings of the 2018 Conference on Human Information Interaction \& Retrieval}} (New Brunswick, NJ, USA) \emph{(\bibinfo{series}{CHIIR '18})}. \bibinfo{publisher}{Association for Computing Machinery}, \bibinfo{address}{New York, NY, USA}, \bibinfo{pages}{32–41}.
\newblock
\showISBNx{9781450349253}
\urldef\tempurl%
\url{https://doi.org/10.1145/3176349.3176387}
\showDOI{\tempurl}


\bibitem[Trippas et~al\mbox{.}(2015a)]%
        {trippas2015resultspresentation}
\bibfield{author}{\bibinfo{person}{Johanne~R. Trippas}, \bibinfo{person}{Damiano Spina}, \bibinfo{person}{Mark Sanderson}, {and} \bibinfo{person}{Lawrence Cavedon}.} \bibinfo{year}{2015}\natexlab{a}.
\newblock \showarticletitle{Results Presentation Methods for a Spoken Conversational Search System}. In \bibinfo{booktitle}{\emph{Proceedings of the First International Workshop on Novel Web Search Interfaces and Systems}} (Melbourne, Australia) \emph{(\bibinfo{series}{NWSearch '15})}. \bibinfo{publisher}{Association for Computing Machinery}, \bibinfo{address}{New York, NY, USA}, \bibinfo{pages}{13–15}.
\newblock
\showISBNx{9781450337892}
\urldef\tempurl%
\url{https://doi.org/10.1145/2810355.2810356}
\showDOI{\tempurl}


\bibitem[Trippas et~al\mbox{.}(2015b)]%
        {trippas2015length}
\bibfield{author}{\bibinfo{person}{Johanne~R. Trippas}, \bibinfo{person}{Damiano Spina}, \bibinfo{person}{Mark Sanderson}, {and} \bibinfo{person}{Lawrence Cavedon}.} \bibinfo{year}{2015}\natexlab{b}.
\newblock \showarticletitle{Towards Understanding the Impact of Length in Web Search Result Summaries over a Speech-only Communication Channel}. In \bibinfo{booktitle}{\emph{Proceedings of the 38th International ACM SIGIR Conference on Research and Development in Information Retrieval}} (Santiago, Chile) \emph{(\bibinfo{series}{SIGIR '15})}. \bibinfo{publisher}{Association for Computing Machinery}, \bibinfo{address}{New York, NY, USA}, \bibinfo{pages}{991–994}.
\newblock
\showISBNx{9781450336215}
\urldef\tempurl%
\url{https://doi.org/10.1145/2766462.2767826}
\showDOI{\tempurl}


\bibitem[Trippas et~al\mbox{.}(2024b)]%
        {trippas2024adapting}
\bibfield{author}{\bibinfo{person}{Johanne~R. Trippas}, \bibinfo{person}{Damiano Spina}, {and} \bibinfo{person}{Falk Scholer}.} \bibinfo{year}{2024}\natexlab{b}.
\newblock \showarticletitle{Adapting Generative Information Retrieval Systems to Users, Tasks, and Scenarios}.
\newblock In \bibinfo{booktitle}{\emph{Information Access in the Era of Generative AI}}, \bibfield{editor}{\bibinfo{person}{Ryen~W. White} {and} \bibinfo{person}{Chirag Shah}} (Eds.). \bibinfo{publisher}{Springer Nature Switzerland AG}, \bibinfo{address}{Cham, Switzerland}.
\newblock


\bibitem[Vtyurina et~al\mbox{.}(2020)]%
        {vtyurina2020mixed}
\bibfield{author}{\bibinfo{person}{Alexandra Vtyurina}, \bibinfo{person}{Charles~LA Clarke}, \bibinfo{person}{Edith Law}, \bibinfo{person}{Johanne~R Trippas}, {and} \bibinfo{person}{Horatiu Bota}.} \bibinfo{year}{2020}\natexlab{}.
\newblock \showarticletitle{A mixed-method analysis of text and audio search interfaces with varying task complexity}. In \bibinfo{booktitle}{\emph{Proceedings of the 2020 ACM SIGIR on International Conference on Theory of Information Retrieval}}. \bibinfo{pages}{61--68}.
\newblock


\bibitem[Wu et~al\mbox{.}(2022)]%
        {Wu2022chainer}
\bibfield{author}{\bibinfo{person}{Tongshuang Wu}, \bibinfo{person}{Ellen Jiang}, \bibinfo{person}{Aaron Donsbach}, \bibinfo{person}{Jeff Gray}, \bibinfo{person}{Alejandra Molina}, \bibinfo{person}{Michael Terry}, {and} \bibinfo{person}{Carrie~J Cai}.} \bibinfo{year}{2022}\natexlab{}.
\newblock \showarticletitle{PromptChainer: Chaining Large Language Model Prompts through Visual Programming}. In \bibinfo{booktitle}{\emph{Extended Abstracts of the 2022 CHI Conference on Human Factors in Computing Systems}} (New Orleans, LA, USA) \emph{(\bibinfo{series}{CHI EA '22})}. \bibinfo{publisher}{Association for Computing Machinery}, \bibinfo{address}{New York, NY, USA}, Article \bibinfo{articleno}{359}, \bibinfo{numpages}{10}~pages.
\newblock
\showISBNx{9781450391566}
\urldef\tempurl%
\url{https://doi.org/10.1145/3491101.3519729}
\showDOI{\tempurl}


\bibitem[Yunkaporta(2023)]%
        {yunkaporta2023right}
\bibfield{author}{\bibinfo{person}{Tyson Yunkaporta}.} \bibinfo{year}{2023}\natexlab{}.
\newblock \bibinfo{booktitle}{\emph{Right story, wrong story: Adventures in Indigenous thinking}}.
\newblock \bibinfo{publisher}{Text Publishing}.
\newblock
\urldef\tempurl%
\url{https://hdl.handle.net/10779/DRO/DU:24502765.v1}
\showURL{%
\tempurl}


\bibitem[Zamani et~al\mbox{.}(2023)]%
        {zamani2023conversational}
\bibfield{author}{\bibinfo{person}{Hamed Zamani}, \bibinfo{person}{Johanne~R Trippas}, \bibinfo{person}{Jeff Dalton}, \bibinfo{person}{Filip Radlinski}, {et~al\mbox{.}}} \bibinfo{year}{2023}\natexlab{}.
\newblock \showarticletitle{Conversational information seeking}.
\newblock \bibinfo{journal}{\emph{Foundations and Trends{\textregistered} in Information Retrieval}} \bibinfo{volume}{17}, \bibinfo{number}{3-4} (\bibinfo{year}{2023}), \bibinfo{pages}{244--456}.
\newblock


\end{thebibliography}


\end{document}